\documentclass[10pt, conference]{IEEEtran}
\IEEEoverridecommandlockouts
\usepackage{cite}
\usepackage{amsmath,amssymb,amsfonts}
\usepackage{algorithm}
\usepackage{algpseudocode}
\usepackage{graphicx}
\usepackage{subfig}
\usepackage{xcolor}
\usepackage{bm}
\usepackage{textcomp}
\usepackage{caption}

\def\BibTeX{{\rm B\kern-.05em{\sc i\kern-.025em b}\kern-.08em
    T\kern-.1667em\lower.7ex\hbox{E}\kern-.125emX}}

\definecolor{darkgreen}{rgb}{0.0, 0.5, 0.0}

\begin{document}

\title{Channel Prediction-Based Physical Layer Authentication under Consecutive Spoofing Attacks}

\author{
\IEEEauthorblockN{
Yijia~Guo\IEEEauthorrefmark{1}\IEEEauthorrefmark{2},
Junqing~Zhang\IEEEauthorrefmark{1}, and
Y.-W. Peter Hong\IEEEauthorrefmark{2}
}

\IEEEauthorblockA{
\IEEEauthorrefmark{1}
School of Computer Science and Informatics, University of Liverpool, Liverpool, L69 3DR, United Kingdom\\ 
\IEEEauthorrefmark{2}
Institute of Communications Engineering,
National Tsing Hua University,
Hsinchu, Taiwan 300044\\ 
Emails: Yijia.Guo@liverpool.ac.uk, Junqing.Zhang@liverpool.ac.uk, ywhong@ee.nthu.edu.tw}

\thanks{The work of J. Zhang was supported in part by the UK Engineering and Physical Sciences Research Council (EPSRC) under grant ID EP/Y037197/1 and in part by Royal Society Research Grants under grant ID RGS/R1/231435.}
}

\maketitle

\begin{abstract}
Wireless networks are highly vulnerable to spoofing attacks, especially when attackers transmit consecutive spoofing packets. Conventional physical layer authentication (PLA) methods have mostly focused on single-packet spoofing attack. However, under consecutive spoofing attacks, they become ineffective due to channel evolution caused by device mobility and channel fading. To address this challenge, we propose a channel prediction-based PLA framework. Specifically, a Transformer-based channel prediction module is employed to predict legitimate CSI measurements during spoofing interval, and the input of channel prediction module is adaptively updated with predicted or observed CSI measurements based on the authentication decision to ensure robustness against sustained spoofing. Simulation results under Rayleigh fading channels demonstrate that the proposed approach achieves low prediction error and significantly higher authentication accuracy than conventional benchmark, maintaining robustness even under extended spoofing attacks.
\end{abstract}


\section{Introduction}
With the rapid development of wireless technologies, device authentication is critical since the broadcast nature of wireless channels makes them susceptible to spoofing attacks~\cite{xie2021survey}. To mitigate such threats, physical layer authentication (PLA) has attracted considerable attention as it leverages the unique channel state information (CSI) that is hard to forge or replicate.

Early studies of PLA predominantly employed statistical modelling, with particular emphasis on hypothesis testing frameworks. Notable methods include Neyman–Pearson detection~\cite{xiao2009channel}, difference-based detection~\cite{xiao2008using, liu2013two, liu2016physical, xie2021physical, zhang2021exploiting}, and Pearson correlation-based detection~\cite{liu2018authenticating}. Although theoretically well-grounded, hypothesis testing methods inherently rely on precise channel statistics, which is often difficult to obtain in real-world scenarios, thereby limiting their practical applicability.

With the advancement of data-driven technologies, a wide range of learning-based approaches have been explored. Initial research employed channel prediction–based authentication with recurrent neural networks, where long short-term memory (LSTM) and gated recurrent units (GRU) were applied to forecast future CSI and enable mean square error (MSE)-based detection~\cite{germain2021channel}. Similarly,~\cite{wang2021channel} proposed a prediction framework that integrates historical CSI with the transmitter’s geographical information. In parallel, similarity learning methods were also explored. A Siamese network based on fully connected networks (FCN) was introduced in~\cite{zhang2025enhancing}. In our previous work~\cite{guo2025practical}, a convolutional neural network (CNN)-based Siamese architecture was proposed to improve robustness. Despite these advances, the existing studies have predominantly focused on scenarios involving a single spoofing packet, leaving the more challenging case of consecutive attack packets largely unexplored.

In practice, attackers may transmit consecutive spoofing packets, which pose two critical challenges. First, the authentication system must reliably detect sustained spoofing during consecutive malicious transmissions. Second, once the legitimate user resumes transmission, the system must avoid false alarms caused by channel decorrelation accumulated in the spoofing interval. Addressing these challenges requires models that can effectively capture channel evolution patterns over prolonged time intervals, so as to distinguish persistent spoofing from natural channel variations.

Recently, Transformer-based channel prediction has gained considerable attention in wireless communications because of its ability to capture long-range temporal correlation. A Transformer-based parallel channel prediction framework was proposed in~\cite{jiang2022accurate} to effectively overcome the error propagation issue. The Transformer architecture is further leveraged in~\cite{zhou2024transformer} for CSI feedback enhancement in AI-native air interfaces, showing superior prediction accuracy on real high-speed railway measurement data. To address vehicular communication challenges, a Transformer-based predictor within a Rate-Splitting Multiple Access (RSMA) enabled V2X framework was introduced in~\cite{zhang2024transformer} to achieve robust interference management and improved throughput. 

Motivated by the strong capability of Transformers in modelling channel evolution, we design a channel prediction-based PLA framework that integrates Transformer-based channel prediction with Pearson correlation-based authentication. In contrast to conventional approaches, our method explicitly predicts the evolution of legitimate CSI during spoofing intervals, thereby facilitating accurate authentication once legitimate transmissions resume. The key contributions of this paper are outlined below.
\begin{itemize}
    \item A Transformer-based channel prediction framework is employed to forecast the evolution of legitimate CSI, where the predicted CSI is utilized to authenticate the observed CSI. Based on the authentication results, the Transformer input is adaptively updated with predicted or the observed CSI measurements, which ensures robustness against sustained spoofing.
    \item Extensive simulations under Rayleigh fading channel validate the proposed scheme, showing significant performance gains over conventional Pearson correlation benchmark in terms of authentication accuracy.
\end{itemize}

\section{System Model and Problem Statement}
\label{sec:system}
\subsection{System Model}
\begin{figure}[!t]
\centerline{\includegraphics[width=3in]{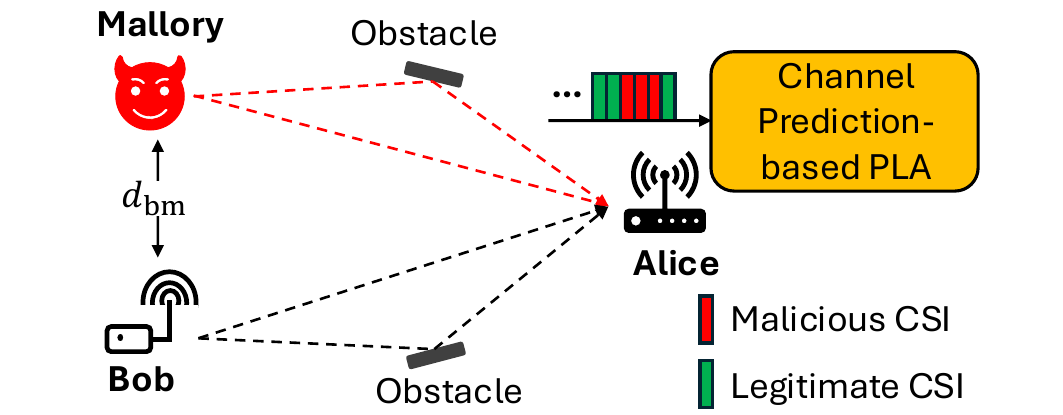}}
\caption{The system model under consecutive spoofing attacks.}
\label{fig:SystemModel}
\end{figure}

As shown in Fig.~\ref{fig:SystemModel}, Bob and Alice communicate over a time-varying multipath channel based on IEEE 802.11n OFDM standard, operating on a 20MHz bandwidth with $M=64$ subcarriers. Meanwhile, an attacker, Mallory, attempts to impersonate Bob by sending consecutive spoofed packets. For each incoming packet, Alice performs channel estimation using the preamble, where $M^{\prime}=52$ subcarriers are active, to extract the CSI, which is then fed into a channel prediction-based PLA scheme to verify its authenticity.

Suppose the $k$-th packet received by Alice is transmitted by Bob. The corresponding channel between Bob and Alice can be represented by
\begin{equation}
\bm{h}_{\rm ba}^{[k]} \triangleq \big[h_{\rm ba}^{[k]}[0], \ldots, h_{\rm ba}^{[k]}[l], \ldots, h_{\rm ba}^{[k]}[L-1]\big],
\end{equation}
where $L$ is the total number of channel taps. Building on the channel model in our previous work~\cite{guo2025practical}, the CSI can be expressed as
\begin{equation}
\label{eq:lsEstimation_vector}
\bm{\widehat{H}}_{\rm ba}^{[k]} =\bm{H}_{\rm ba}^{[k]}+\bm{\widehat{Z}},
\end{equation}
where $\bm{H}_{\rm ba}^{[k]} \in \mathbb{C}^{M^{\prime} \times 1}$ is the channel frequency response (CFR) and $\bm{\widehat{Z}}$ is the estimation noise.

\textbf{Threat Model:}
This paper considers a scenario in which the attacker, Mallory, located at a distance $d_{\rm bm}$ from Bob, successfully spoofs Bob’s MAC address and transmits malicious packets to Alice. Equipped with knowledge of the wireless protocol employed by Alice and Bob, e.g., channel bandwidth, carrier frequency, modulation format, and OFDM frame structure, Mallory can maintain the impersonation across multiple transmissions. By persistently injecting consecutive spoofed packets to Alice, Mallory poses a sustained and severe threat to the physical layer security of the system.

\subsection{Problem Statement}
Earlier studies have mostly focused on single-packet spoofing attack. However, in practical wireless environments, attackers can send consecutive spoofed packets to maintain a persistent impersonation. In such cases, Alice typically authenticates the $(k+n)$-th received packet by comparing its observed CSI $\widehat{\bm{H}}^{[k+n]}$ with the reference CSI measurement $\widehat{\bm{H}}_{\rm ba}^{[k]}$ obtained from Bob’s most recent legitimate transmission. This method works only when the channel between Bob and Alice changes slowly or the spoofing duration is short, so that the reference CSI measurement $\widehat{\bm{H}}_{\rm ba}^{[k]}$ can still represent the actual channel conditions.

When Mallory conducts long spoofing attacks, the legitimate channel gradually changes due to user mobility and channel fading, making $\widehat{\bm{H}}_{\rm ba}^{[k]}$ outdated as $n$ increases. The difference between the reference CSI and the current CSI then grows, causing Alice to incorrectly reject Bob’s subsequent legitimate packets. Therefore, extended spoofing significantly reduces authentication accuracy and reveals a key limitation of traditional schemes.

% Existing research has largely addressed scenarios involving a single spoofing packet, but such approaches are no longer applicable in the case of continuous spoofing attacks. Specifically, when Alice receives the $(k+N_{a}+1)$-th packet from Bob, she typically compares the observed $\widehat{\bm{H}}^{[k+N_{a}+1]}$ with the reference CSI measurement $\widehat{\bm{H}}_{\rm ba}^{[k]}$ obtained from Bob’s most recent legitimate packet to determine the authenticity. This strategy is effective only if the channel remains stable during the attack transmission, or if the attack period is short enough for the newly observed $\widehat{\bm{H}}^{[k+N_{a}+1]}$ to remain highly correlated with the last trusted measurement $\widehat{\bm{H}}_{\rm ba}^{[k]}$. However, if Mallory launches a prolonged spoofing transmission, the channel between Bob and Alice evolves naturally due to mobility and fading. Consequently, by the time Bob transmits his next legitimate packet, the conventional methods may mistakenly classify Bob’s new packet as suspicious since the observed $\widehat{\bm{H}}^{[k+N_{a}+1]}$ may have decorrelated from $\widehat{\bm{H}}_{\rm ba}^{[k]}$. Therefore, long attack intervals significantly increase detection difficulty and expose a fundamental limitation of conventional authentication schemes.

\section{Proposed Method}
\label{sec:proposed}
We propose a channel prediction-based framework that leverages Transformer-based channel prediction for robust authentication. This section introduces the overall framework, followed by the details of the training stage and test stage.

\subsection{Overall Framework}
\begin{figure}[!t]
  \centering
  \includegraphics[width=3.2in]{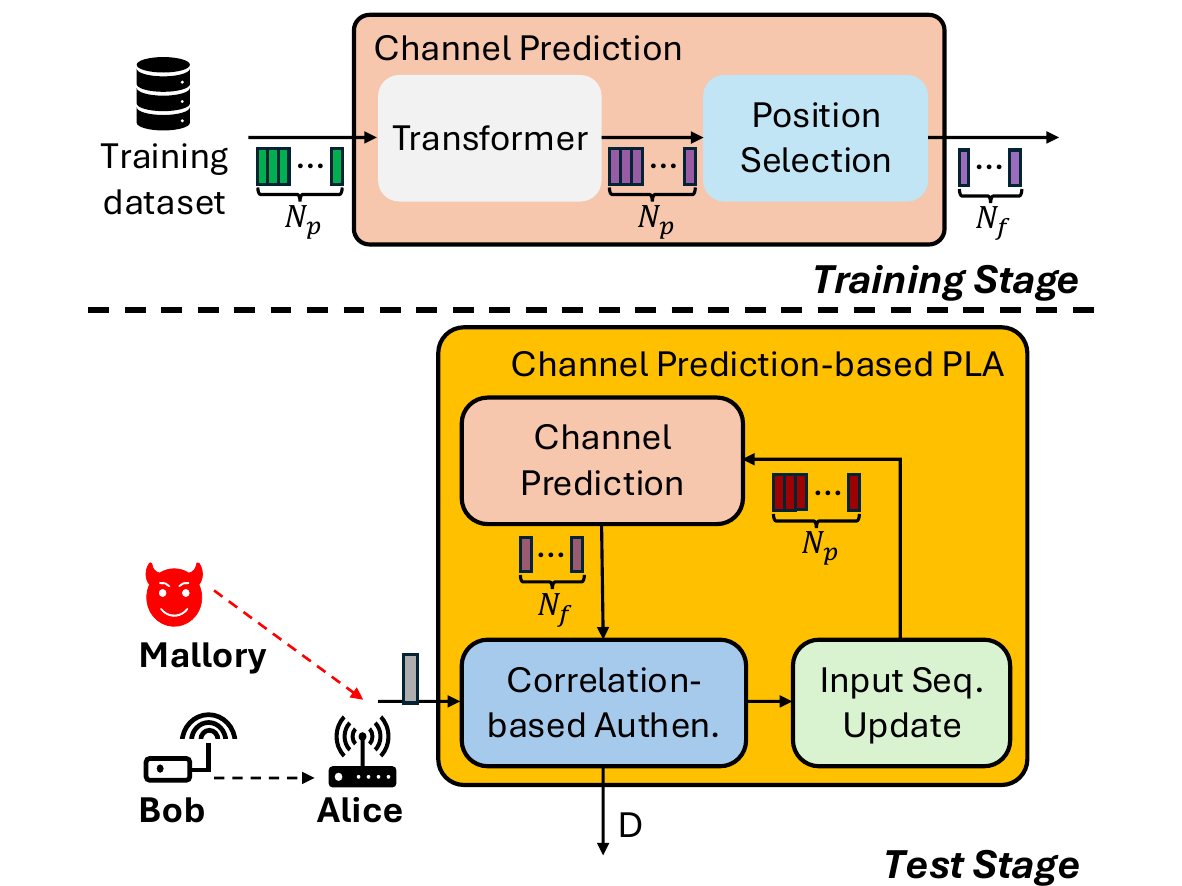}
  \caption{The channel prediction-based PLA system.}
  \label{fig:AuthenticationSystem}
\end{figure}

As illustrated in Fig.~\ref{fig:AuthenticationSystem}, the proposed channel prediction-based PLA system operates in a training stage and a test stage. 
In the training stage, a channel prediction module, which is composed of a Transformer and a position selection module, is trained to predict $N_{f}$ future CSI measurements based on $N_{p}$ input CSI measurements, thereby learning to capture the temporal evolution of the wireless channel. The details of the training stage will be introduced in Section~\ref{subsec:training}. 
During the test stage, the channel prediction-based PLA system operates in an iterative manner. In each iteration, the trained channel prediction module obtains $N_{f}$ future CSI measurements, which are utilised by correlation-based authentication module to conduct authentication of the incoming packets. Subsequently, the input sequence of the channel prediction module is updated based on the authentication decision. Further details of the test stage are provided in Section~\ref{subsec:test}. 

Each CSI measurement, complex values collected over $M^{\prime}$ subcarriers, are preprocessed into a real-valued representation by concatenating its real and imaginary parts, defined as
\begin{equation}
\widehat{\bm{\mathbb{H}}}^{[k]} \triangleq \big[\Re\{\widehat{\bm H}^{[k]}\}^{T}, \Im\{\widehat{\bm H}^{[k]}\}^{T}\big]^{T} \in \mathbb{R}^{2M^{\prime}}.
\end{equation}

\subsection{Training Stage}
\label{subsec:training}
\subsubsection{Transformer}
\begin{figure}[!t]
  \centering
  \includegraphics[width=3.2in]{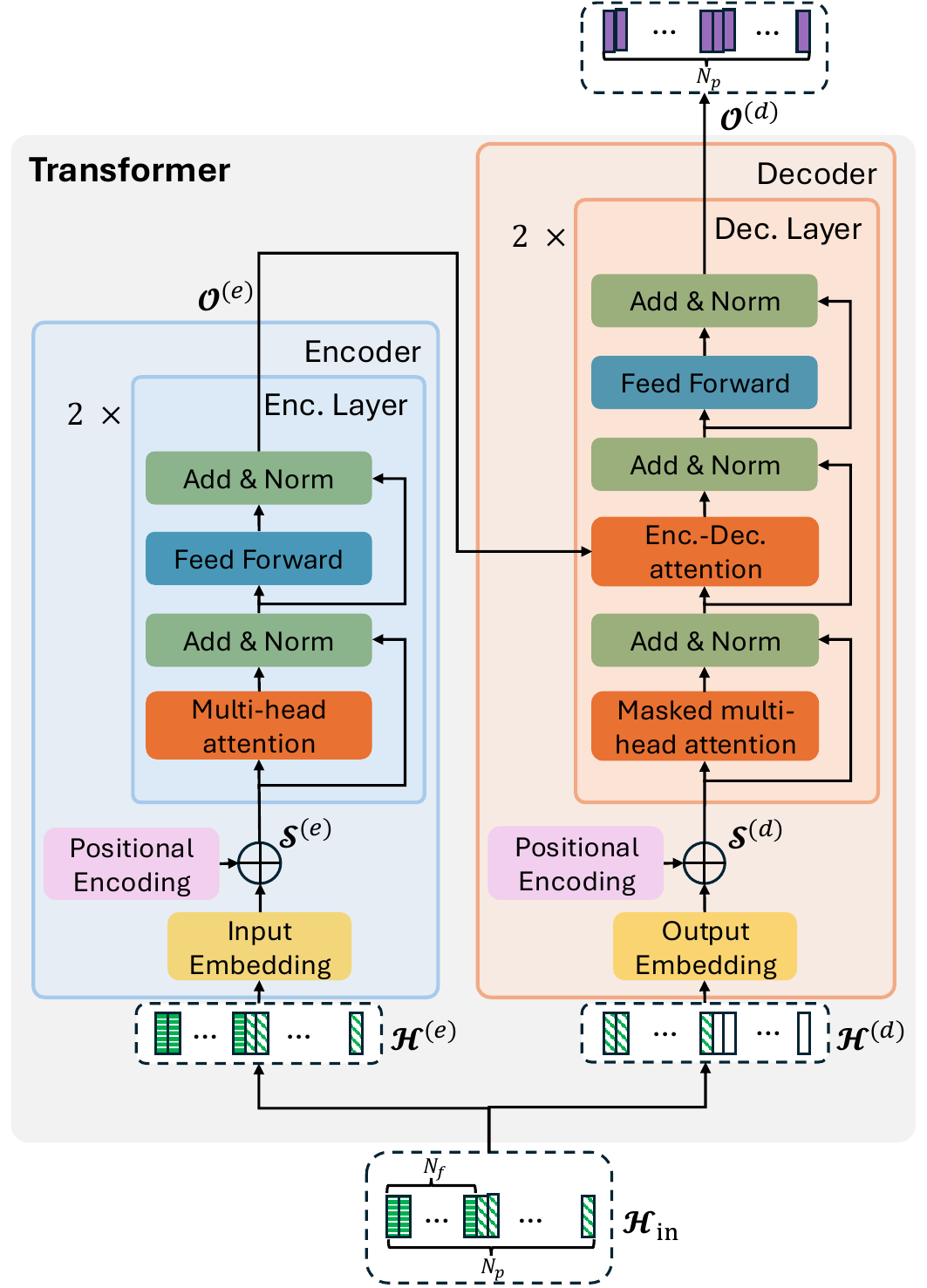}
  \caption{The structure of encoder-decoder Transformer.}
  \label{fig:Transformer}
\end{figure}

As shown in Fig.~\ref{fig:Transformer}, the Transformer framework comprises an encoder that extracts temporal features from historical CSI and a decoder that predicts upcoming CSI. In this paper, both the encoder and the decoder are composed of two identical layers. For illustrative purposes, we describe the structure of a single encoder layer and a single decoder layer.

The raw input sequence to the Transformer consists of $N_{p}$ CSI measurements, given as
\begin{equation}
    \bm{\mathcal{H}}_{\rm in} \triangleq \big[\widehat{\bm{\mathbb{H}}}^{[1]}, \ldots, \widehat{\bm{\mathbb{H}}}^{[N_{p}]} \big] \in \mathbb{R}^{2M^{\prime} \times N_{p}},
\end{equation} 
which is used to build the input for encoder and decoder.

\textbf{Encoder:}
The encoder input sequence, denoted as 
\begin{equation}
    \bm{\mathcal{H}}^{(e)} \triangleq \bm{\mathcal{H}}_{\rm in} = \big[\widehat{\bm{\mathbb{H}}}^{[1]}, \ldots, \widehat{\bm{\mathbb{H}}}^{[N_{p}]} \big] \in \mathbb{R}^{2M^{\prime} \times N_{p}}, 
\end{equation} 
takes the same $N_{p}$ CSI measurements as input. After input embedding and positional encoding, the encoder layer input is denoted as $\bm{\mathcal{S}}^{(e)} \in \mathbb{R}^{N_{p} \times d_{m}}$, where $d_{m}$ represents the embedding dimension.

In the encoder layer, the multi-head attention mechanism projects $\bm{\mathcal{S}}^{(e)}$ into queries, keys and values using $N_{\rm head}$ groups of learnable matrices. Specifically, the $i$-th head computes
\begin{equation}
    \bm{X}_{i}^{(e)} = f_{\text{sm}}\Big(\frac{\bm{\mathcal{S}}^{(e)} \bm{W}_{Q_{i}}^{(e)} \cdot (\bm{\mathcal{S}}^{(e)} \bm{W}_{K_{i}}^{(e)})^T}{\sqrt{d_{k}}}\Big) \cdot \bm{\mathcal{S}}^{(e)} \bm{W}_{V_{i}}^{(e)},
\end{equation}
where $\bm{W}_{Q_{i}}^{(e)}, \bm{W}_{K_{i}}^{(e)}, \bm{W}_{V_{i}}^{(e)} \in \mathbb{R}^{d_{m}\times d_{k}}$, with $d_{k}=d_{m}/N_{\rm head}$, and $f_{\text{sm}}(\cdot)$ denotes the softmax function. The outputs of all heads are then concatenated and linearly projected, given as
\begin{equation}
\bm{X}^{(e)}=\text{concat}\big(\bm{X}_{1}^{(e)},\ldots, \bm{X}_{i}^{(e)}, \ldots, \bm{X}_{N_{\rm head}}^{(e)}\big) \cdot \bm{W}_{O}^{(e)},
\end{equation}
with $\bm{W}_{O}^{(e)} \in \mathbb{R}^{d_{m} \times d_{m}}$ is the output projection. 
The attention output is subsequently combined with the original input via a residual connection and then normalized by layer normalization, expressed as
\begin{equation}
    \bm{Z}^{(e)} = f_{\rm LN}\big(\bm{\mathcal{S}}^{(e)} + \bm{X}^{(e)}\big),
\end{equation}
where $f_{\rm LN}(\cdot)$ denotes layer normalization. The normalized features are further refined by a feed-forward network, followed by an residual connection and layer normalization, given as
\begin{equation}
    \bm{O}^{(e)} = f_{\rm LN}\big(\bm{Z}^{(e)} + f_{\rm FFN}(\bm{Z}^{(e)})\big),
\end{equation}
where $f_{\rm FFN}(\cdot)$ denotes the feed-forward network. The output $\bm{O}^{(e)}$ represents the encoded historical CSI features.

\textbf{Decoder:}
The decoder input sequence, denoted as 
\begin{equation}
    \bm{\mathcal{H}}^{(d)} \triangleq \big[\widehat{\bm{\mathbb{H}}}^{[N_{f}+1]}, \ldots, \widehat{\bm{\mathbb{H}}}^{[N_{p}]}, \bm{0}_{2M^{\prime} \times N_{f}} \big] \in \mathbb{R}^{2M^{\prime} \times N_{p}}, 
\end{equation} 
is initialized with the latest $(N_{p}-N_{f})$ samples in $\bm{\mathcal{H}}_{\rm in}$ and zero-padded for the remaining $N_{f}$ positions, where $N_{f} < N_{p}$. After output embedding and positional encoding, the decoder layer input is denoted as $\bm{\mathcal{S}}^{(d)} \in \mathbb{R}^{N_{p} \times d_{m}}$.

In the decoder layer, the masked multi-head attention mechanism projects $\bm{\mathcal{S}}^{(d)}$ into queries, keys and values through $N_{\rm head}$ groups of learnable matrices. Specifically, the $i$-th head computes
\begin{equation}
    \bm{X}_{i}^{(d)} = f_{\text{sm}}\Big(\frac{\bm{\mathcal{S}}^{(d)} \bm{W}_{Q_{i}}^{(d)} \cdot (\bm{\mathcal{S}}^{(d)} \bm{W}_{K_{i}}^{(d)})^T}{\sqrt{d_{k}}}+\bm{M}^{(d)}\Big) \cdot
    \bm{\mathcal{S}}^{(d)} \bm{W}_{V_{i}}^{(d)},
\end{equation}
where $\bm{W}_{Q_{i}}^{(d)}, \bm{W}_{K_{i}}^{(d)}, \bm{W}_{V_{i}}^{(d)} \in \mathbb{R}^{d_{m}\times d_{k}}$, and $\bm{M}^{(d)} \in \mathbb{R}^{N_{p} \times N_{p}}$ is a casual mask applied to prevent access to future positions, defined as
\begin{equation}
M^{(d)}[i,j] =
\begin{cases}
0, & j \leq i; \\
-\infty, & j > i,
\end{cases}
\quad i, j \in \{1,2,\ldots,N_{p}\}.
\end{equation}
The outputs of all heads are then concatenated and linearly projected, given as
\begin{equation}
\bm{X}^{(d)}=\text{concat}\big(\bm{X}_{1}^{(d)},\ldots, \bm{X}_{i}^{(d)}, \ldots, \bm{X}_{N_{\rm head}}^{(d)}\big) \cdot \bm{W}_{O}^{(d)},
\end{equation}
with $\bm{W}_{O}^{(d)} \in \mathbb{R}^{d_{m} \times d_{m}}$ is the output projection. The attention output is then processed via a residual connection and layer normalization, given as
\begin{equation}
    \bm{Z}^{(d)} = f_{\rm LN}\big(\bm{\mathcal{S}}^{(d)} + \bm{X}^{(d)}\big).
\end{equation}
Next, the encoder–decoder attention is further performed using $\bm{Z}^{(d)}$ as queries, and $\bm{O}^{(e)}$ as keys and values, expressed as
\begin{equation}
    \bm{X}_{i}^{(c)} = f_{\text{sm}}\Big(\frac{\bm{Z}^{(d)} \bm{W}_{Q_{i}}^{(c)} \cdot (\bm{O}^{(e)} \bm{W}_{K_{i}}^{(c)})^T}{\sqrt{d_{k}}}\Big) \cdot \bm{O}^{(e)} \bm{W}_{V_{i}}^{(c)}.
\end{equation}
Then the outputs of all heads are concatenated and linearly projected, given as
\begin{equation}
    \bm{X}^{(c)}=\text{concat}\big(\bm{X}_{1}^{(c)},\ldots, \bm{X}_{i}^{(c)}, \ldots, \bm{X}_{N_{\rm head}}^{(c)}\big) \cdot \bm{W}_{O}^{(c)},
\end{equation}
where $\bm{W}_{O}^{(c)} \in \mathbb{R}^{d_{m} \times d_{m}}$ is an output projection. The attention output is then passed to a residual connection and layer normalisation, given as
\begin{equation}
    \bm{Z}^{(c)} = f_{\rm LN}\big(\bm{Z}^{(d)} + \bm{X}^{(c)}\big).
\end{equation}
Finally, the predicted CSI is obtained after feed-forward network, followed by an additional residual connection and layer normalisation, expressed as
\begin{equation}
    \bm{O}^{(d)} = f_{\rm LN}\big(\bm{Z}^{(c)} + f_{\rm FFN}(\bm{Z}^{(c)})\big).
\end{equation}

\subsubsection{Position Selection}
The decoder output can be explicitly expressed as 
\begin{equation}
    \bm{O}^{(d)}=\big[\widetilde{\bm{\mathbb{H}}}^{[N_{f}+1]}, \ldots, \widetilde{\bm{\mathbb{H}}}^{[N_{p}]}, \widetilde{\bm{\mathbb{H}}}^{[N_{p}+1]}, \ldots, \widetilde{\bm{\mathbb{H}}}^{[N_{p}+N_{f}]}\big]^{T}, 
\end{equation}
where only the last $N_{f}$ rows correspond to the predicted CSI for the next $N_{f}$ packets. Therefore, the predicted CSI sequence is given as
\begin{equation}
    \bm{\mathcal{H}}_{\text{pred}}= f_{\rm sel}(\bm{O}^{(d)})\triangleq \big[\widetilde{\bm{\mathbb{H}}}^{[N_{p}+1]}, \ldots, \widetilde{\bm{\mathbb{H}}}^{[N_{p}+N_{f}]}\big].
\end{equation}

The transformer is then trained by minimizing the normalized mean squared error (NMSE) between the predicted CSI and the corresponding ground truth, defined as 
\begin{equation}
\mathcal{L} = \mathbb{E}\left\{ 
\frac{\sum_{n=1}^{N_{f}} \Big\|\widetilde{\bm{\mathbb{H}}}^{[N_{p}+n]} - \widehat{\bm{\mathbb{H}}}^{[N_{p}+n]}\Big\|^{2}}{\sum_{n=1}^{N_{f}} \Big\|\widehat{\bm{\mathbb{H}}}^{[N_{p}+n]}\Big\|^{2}} 
\right\}.
\end{equation}

\subsection{Test Stage}
\label{subsec:test}
In the test stage, the channel prediction-based PLA system operates in an iterative manner. Firstly, we assume that the historical $N_{p}$ received packets are from legitimate user Bob. The corresponding CSI sequence is denoted as 
\begin{equation}
    \bm{\mathcal{H}}_{\text{hist}} =[\widehat{\bm{\mathbb{H}}}_{\rm ba}^{[k-N_{p}+1]}, \ldots, \widehat{\bm{\mathbb{H}}}_{\rm ba}^{[k]}].
\end{equation}
which is used as the initial input to the channel prediction module to obtain an initial and reliable channel prediction. Furthermore, there exists an observed CSI sequence requiring authentication, represented as
\begin{equation}
    \label{eq:rxSeq}
    \bm{\mathcal{H}}_{\text{rx}}^{\prime} = [\widehat{\bm{\mathbb{H}}}^{[k+1]}, \widehat{\bm{\mathbb{H}}}^{[k+2]}, \ldots].
\end{equation}

% Given the input length $N_{p}$, the prediction length $N_{f}$, trained Transformer model $\mathcal{M}(\cdot)$, historical CSI sequence $\bm{\mathcal{H}}_{\text{hist}}$, and observed CSI sequence to be authenticated $\bm{\mathcal{H}}_{\text{rx}}^{\prime}$ as inputs, the proposed channel prediction-based PLA scheme is summarized in Algorithm~\ref{alg:Pred_Authen}.
Let $\bm{\mathcal{H}}_{\text{in}}^{i}$ denote the input sequence fed into channel prediction module at the $i$-th iteration, with $\bm{\mathcal{H}}_{\text{in}}^{1}=\bm{\mathcal{H}}_{\text{hist}}$, and let $N_{\text{auth}}^{i}$ denote the number of packets authenticated before the $i$-th iteration, with $N_{\text{auth}}^{1}=0$. The proposed channel prediction-based PLA scheme is summarized in Algorithm~\ref{alg:Pred_Authen}.
\begin{algorithm}[!h]
\caption{Channel Prediction-based Authentication}
\label{alg:Pred_Authen}
\begin{algorithmic}[1]
\Require Input length $N_{p}$, prediction length $N_{f}$, trained Transformer model $\mathcal{M}(\cdot)$, historical legitimate CSI sequence $\bm{\mathcal{H}}_{\text{hist}}$, observed CSI sequence to be authenticated $\bm{\mathcal{H}}_{\text{rx}}^{\prime}$;
\Ensure $\bm{D}$;

\State {\bf Initialization:} $\bm{\mathcal{H}}_{\text{in}}^{1} \gets \bm{\mathcal{H}}_{\text{hist}}$, $N_{\text{auth}}^{1} \gets 0$, $i \gets 1$;

\While{true}
    \Statex \text{\hspace{0.5cm}\% $i$-th iteration}
    \State Generate $\bm{\mathcal{H}}_{\text{pred}}^{i}$ and $\bm{\mathcal{H}}_{\text{rx}}^{i}$ based on~\eqref{eq:Test_H_Gen};
    \For{$j=1$ to $N_{f}$}
        \State Calculate $r_{i,j}$ based on~\eqref{eq:Test_corr_cal};
        \If{$r_{i,j} \geq \epsilon$}
            \State $\bm{\mathcal{H}}_{\text{upd}}^{i}[j] = \bm{\mathcal{H}}_{\text{rx}}^{i}[j]$;
            \State $D[N_{\text{auth}}^{i}+j] = 1$;
            \State break; \label{alg:Test_break}
        \Else
            \State $\bm{\mathcal{H}}_{\text{upd}}^{i}[j] = \bm{\mathcal{H}}_{\text{pred}}^{i}[j]$;
            \State $D[N_{\text{auth}}^{i}+j] = 0$;
        \EndIf
    \EndFor \label{alg:Test_innerLoop_end}
    \State $N_{\text{auth}}^{i+1} \gets N_{\text{auth}}^{i}+j$;
    \State Generate $\bm{\mathcal{H}}_{\text{in}}^{i+1}$ using $\bm{\mathcal{H}}_{\text{in}}^{i}$ and $\bm{\mathcal{H}}_{\text{upd}}^{i}$ based on~\eqref{eq:input_seq_upd};
    \State $i \gets i+1$;
\EndWhile
\end{algorithmic}
\end{algorithm}

\subsubsection{Channel Prediction}
In $i$-th iteration, the channel prediction sequence $\bm{\mathcal{H}}_{\text{pred}}^{i}$ and the incoming sequence that needs to be authenticated are extracted as
\begin{subequations}
\label{eq:Test_H_Gen}
\begin{align}
    \bm{\mathcal{H}}_{\text{pred}}^{i} &= f_{\text{sel}}\big(\mathcal{M}_{\rm T}(\bm{\mathcal{H}}_{\text{in}}^{i})\big);\\
    \label{eq:rx_extract}
    \bm{\mathcal{H}}_{\text{rx}}^{i} &= \bm{\mathcal{H}}_{\text{rx}}^{\prime}[N_{\text{auth}}+1: N_{\text{auth}} + N_{f}],
\end{align}
\end{subequations}
where~\eqref{eq:rx_extract} represents extracting the $(N_{\text{auth}}+1)$-th to $(N_{\text{auth}}+N_{f})$-th columns of $\bm{\mathcal{H}}_{\text{rx}}^{\prime}$.

\subsubsection{Correlation-based Authentication}
Authentication is conducted progressively on each prediction step $j=1,\dots,N_{f}$. In $i$-th iteration, for each $j$-th prediction step, the Pearson correlation between predicted CSI $\bm{\mathcal{H}}_{\text{pred}}^{i}[j]$ and the incoming CSI $\bm{\mathcal{H}}_{\text{rx}}^{i}[j]$ is calculated, given as
\begin{equation}
\label{eq:Test_corr_cal}
    r_{i,j} = f_{\text{corr}}(\bm{\mathcal{H}}_{\text{pred}}^{i}[j], \bm{\mathcal{H}}_{\text{rx}}^{i}[j]),
\end{equation}
where $f_{\text{corr}}(\cdot,\cdot)$ is defined as the Pearson correlation function. After obtaining the correlation $r_{i,j}$, the authentication decision can be made based on
\begin{equation}
D[N_{\text{auth}}^{i}+j]=
\begin{cases}
    1, & \text{if } r_{i,j} \geq \epsilon;\\
    0, & \text{if } r_{i,j} < \epsilon,
\end{cases}
\end{equation}
where $D[N_{\text{auth}}^{i}+j]=1$ indicates a legitimate packet, $D[N_{\text{auth}}^{i}+j]=0$ denotes a malicious packet, and the threshold $\epsilon$ is obtained empirically through experiments.

\subsubsection{Input Sequence Update}
The update sequence $\bm{\mathcal{H}}_{\text{upd}}^{i}$ is constructed adaptively based on the authentication decision, 
\begin{equation}
\bm{\mathcal{H}}_{\text{upd}}^{i}[j] = 
\begin{cases}
    \bm{\mathcal{H}}_{\text{rx}}^{i}[j], & \text{if } D[N_{\text{auth}}^{i}+j]=1; \\
	\bm{\mathcal{H}}_{\text{pred}}^{i}[j], & \text{if } D[N_{\text{auth}}^{i}+j]=0.
\end{cases}
\end{equation}
For legitimate packets (i.e., $D[N_{\text{auth}}^{i}+j]=1$), the observed CSI is included to enhance prediction fidelity, whereas for malicious packets (i.e., $D[N_{\text{auth}}^{i}+j]=0$), the predicted CSI is incorporated to mitigate the impact of adversarial measurements. By adaptively choosing between the predicted and observed CSI according to the decision, this update strategy ensures that the Transformer is always guided by reliable CSI inputs, thereby maintaining stable and resilient prediction performance under both normal and attack conditions.

Once either a legitimate packet has been detected (line~\ref{alg:Test_break}) or all the $N_{f}$ predicted CSI in $i$-th iteration have been consumed (line~\ref{alg:Test_innerLoop_end}), $N_{\text{auth}}^{i}$ is updated by accumulating the number of CSI measurements authenticated during the $i$-th iteration and the input sequence of the channel prediction module is updated according to
\begin{equation}
\label{eq:input_seq_upd}
    \bm{\mathcal{H}}_{\text{in}}^{i+1} = \text{concat} (\bm{\mathcal{H}}_{\text{in}}^{i}[j+1:N_{p}], \bm{\mathcal{H}}_{\text{upd}}^{i}).
\end{equation}
The iteration index $i$ is incremented by 1. Then a new prediction iteration is triggered with the updated input sequence $\bm{\mathcal{H}}_{\text{in}}^{i+1}$. The design of triggering a new iteration upon the detection of a legitimate packet ensures that the Transformer always exploits the maximum number of the most recent observed legitimate CSI measurements at each prediction step, thereby improving accuracy.

\section{Performance Metrics and Benchmark}
In this section, we first introduce the performance metrics adopted to evaluate both the channel prediction quality and the authentication reliability of the proposed scheme. Subsequently, we describe the benchmark authentication method.

\subsection{Performance Metrics}
To evaluate the effectiveness of the proposed channel prediction-based authentication scheme, we adopt two performance metrics.

\subsubsection{CSI Prediction Error}
The quality of channel prediction is assessed by NMSE between the predicted and ground truth CSI measurements. In contrast to the training stage, where the loss is defined as the averaged NMSE across $N_{f}$ prediction steps, the evaluation is conducted on a step-wise basis to capture the prediction quality at each prediction step.

\subsubsection{Authentication Accuracy}
The authentication performance is measured by the sequence-level accuracy across multiple test sequences with varying spoofing attack lengths. Suppose a collection of test sequences 
% $\{\widehat{\bm{\mathcal{H}}}_{\text{auth}}^{(s)}\}_{s=1}^{N_{s}}$ 
is constructed, each containing a mix of legitimate and malicious packets with different attack lengths. Let $N_{\text{acc}}$ denote the number of sequences in which all packets are authenticated correctly (i.e., all legitimate packets accepted and all malicious packets rejected), and $N_{s}$ be the total number of sequences. The authentication accuracy is then defined as
\begin{equation}
    \text{Accuracy} = \frac{N_{\text{acc}}}{N_{s}}.
\end{equation}
This metric captures the overall reliability of the authentication scheme under practical attack scenarios.

\subsection{Benchmark}
For comparison, we implement a benchmark authentication scheme following the conventional Pearson correlation–based method without channel prediction. In this approach, the receiver uses Pearson correlation to compare each newly observed CSI with the most recent trusted legitimate CSI, which serves as the reference. If the correlation falls below a predefined threshold $\epsilon$, the packet is marked as suspicious; otherwise, it is accepted as legitimate and the newly observed CSI is used to update the reference. Although computationally efficient, this benchmark is susceptible to channel decorrelation during prolonged spoofing attack.

% \begin{algorithm}[t]
% \caption{Benchmark Authentication without Prediction}
% \label{alg:noPred_Authen}
% \begin{algorithmic}[1]
% \Require Preprocessed CSI sequence
% to be authenticated $\widehat{\bm{\mathcal{H}}}_{\text{auth}}=\big[\widehat{\bm{\mathcal{H}}}^{[k+1]}, \widehat{\bm{\mathcal{H}}}^{[k+2]}, \ldots, \widehat{\bm{\mathcal{H}}}^{[k+N]}\big]$;
% \Ensure $\bm{y}_{\text{pred}}$;

% \State {\bf Initialization:} $\bm{\mathcal{H}}_{\rm ref} \gets \widehat{\bm{\mathcal{H}}}_{\rm{ba}}^{[k]}$, $\bm{y}_{\text{pred}} \gets \bm{0}_{1\times N}$;

% \For{$j=1$ to $N$}
%     \If{$\text{corr}(\widehat{\bm{\mathcal{H}}}^{[k+j]}, \bm{\mathcal{H}}_{\rm ref})<\epsilon$}
%         \State $y_{\text{pred}}[j]=1$;
%     \Else
%         \State $\bm{\mathcal{H}}_{\rm ref} = \widehat{\bm{\mathcal{H}}}^{[k+j]}$;
%         \State $y_{\text{pred}}[j]=0$;
%     \EndIf
% \EndFor
% \end{algorithmic}
% \end{algorithm}

\section{Simulation Evaluation}
\label{sec:evaluation}
This section evaluates the channel prediction-based authentication framework under simulated wireless channels. We first describe the simulation setup, then elaborate channel prediction performance, followed by authentication performance.

\subsection{Simulation Setup}
% The dataset is generated using the MATLAB Communications Toolbox\footnote{https://uk.mathworks.com/help/comm/ref/comm.rayleighchannel-system-object.html}. Specifically, a WLAN Non-HT format waveform is passed through a Rayleigh fading channel. Additive white Gaussian noise (AWGN) is then added. At the receiver, the LTS is extracted, and LS estimation is applied to obtain CSI measurements.

\subsubsection{Training Dataset}
The Transformer training dataset is constructed as $\mathcal{D}_{\text{train}}=\{({\bm u}_{i},{\bm v}_{i})\}_{i=1}^{N_{\text{train}}}$, with
\begin{subequations}\label{eq:TrainDataset}
\begin{align}
{\bm u}_{i} & \triangleq[\widehat{\bm{\mathbb{H}}}_{\rm ba}^{[k-N_{p}+1]}, \ldots, \widehat{\bm{\mathbb{H}}}_{\rm ba}^{[k]}];\\
{\bm v}_{i} & \triangleq[\widehat{\bm{\mathbb{H}}}_{\rm ba}^{[k+1]}, \ldots, \widehat{\bm{\mathbb{H}}}_{\rm ba}^{[k+N_{f}]}],
\end{align}
\end{subequations}
where $N_{\text{train}}$ denotes the training dataset size.
The simulation parameter configuration is given as
\begin{itemize}
    \item Channel model: an exponential power delay profile (PDP) with root-mean-square (RMS) delay $T_{\rm rms} \in \{20, 40, 80, 160, 220\}$ ns.
    \item $\text{SNR} \in \{5,8,10,12,13,14,15,16,17,18,19,20,50\}$ dB.
    \item Terminal speed: $v_0 \sim \mathcal{U}(0.5, 2)$ m/s.
    \item Transmission interval: $\Delta t = 3$ ms.
\end{itemize}
The training dataset is generated by iterating over different parameter combinations.

\subsubsection{Test Dataset}
The test sequences $\bm{\mathcal{H}}_{\text{rx}}^{\prime}$ are generated based on
\begin{equation}
    \bm{\mathcal{H}}_{\text{rx}}^{\prime} =\big[ \widehat{\bm{\mathbb{H}}}_{\rm ma}^{[k+1]}, \ldots, \widehat{\bm{\mathbb{H}}}_{\rm ma}^{[k+N_{a}]}, \widehat{\bm{\mathbb{H}}}_{\rm ba}^{[k+N_{a}+1]}, \ldots \big],
\end{equation}
where the first $N_{a}$ CSI samples are malicious, and the remaining samples originate from legitimate packets, with $N_{a}$ denoting the attack length.
% generated based on~\eqref{eq:rxSeq}, where a spoofing block of length $N_{a}$ is inserted at a random position, that is
% \begin{equation}
%     \widehat{\bm{\mathbb{H}}}^{[k+n]} = 
%     \begin{cases}
%         \widehat{\bm{\mathbb{H}}}_{\rm ma}^{[k+n]}, & n \in [n_{s}, n_{s}+N_{a}-1], \\
%         \widehat{\bm{\mathbb{H}}}_{\rm ba}^{[k+n]}, & \text{otherwise},
%     \end{cases}
% \end{equation}
% where $n_{s}$ denotes a random starting point of the attack.

% In addition, the ground truth CSI sequences are generated following
% \begin{equation}
%     \widehat{\bm{\mathcal{H}}}_{\text{legit}} \triangleq [\widehat{\bm{\mathcal{H}}}_{\rm ba}^{[k-N_{p}+1]}, \ldots, \widehat{\bm{\mathcal{H}}}_{\rm ba}^{[k]}, \widehat{\bm{\mathcal{H}}}_{\rm ba}^{[k+1]}, \ldots, \widehat{\bm{\mathcal{H}}}_{\rm ba}^{[k+N]}],
% \end{equation}
% which is used to evaluate the prediction error. 

\subsection{Simulation Results}
\begin{figure}[t]
  \centering
  \includegraphics[width=3.4in]{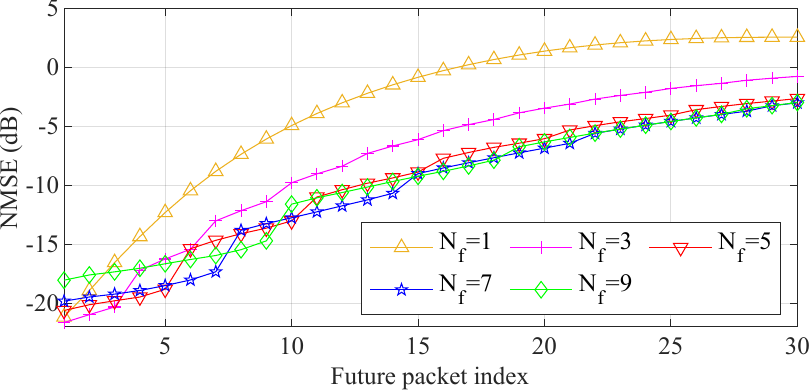}
  \caption{The NMSE of Transformer-based channel prediction module with $N_{\rm head}=4$ and $N_{p}=20$ evaluated on the simulation test dataset under $T_{\rm rms}=50$ ns, $\text{SNR}=20$~dB, $v_0=0.5$ m/s, and $\Delta t=3$ ms.}
  \label{fig:NMSE_predLen}
\end{figure}

In Fig.~\ref{fig:NMSE_predLen}, the prediction error is evaluated under a worst-case assumption where, in Algorithm~\ref{alg:Pred_Authen}, $\bm{\mathcal{H}}_{\text{upd}}^{i}$ is always appended with the predicted CSI $\bm{\mathcal{H}}_{\text{pred}}^{i}[j]$ instead of the observed CSI measurement $\bm{\mathcal{H}}_{\text{rx}}^{i}[j]$. This setting emphasizes the effect of accumulative prediction errors across multiple prediction iterations. Error accumulation occurs only when a new iteration is triggered, which corresponds to the turning points shown in Fig.~\ref{fig:NMSE_predLen}. It can be observed that the prediction errors accumulate most rapidly when $N_{f}=1$. In contrast, for larger prediction lengths (i.e., $N_{f}=3,5,7,9$), the errors grow more gradually. This is because in each new iteration, $N_{f}$ predicted CSI samples from the previous iteration, which already contain prediction errors in themselves, are appended to the input sequence of the channel prediction module, leading to error accumulation. Considering the trade-off between the performance in each iteration and the error accumulation, $N_{f}=5$ offers a favourable balance.

\begin{figure}[t]
  \centering
  \includegraphics[width=3.4in]{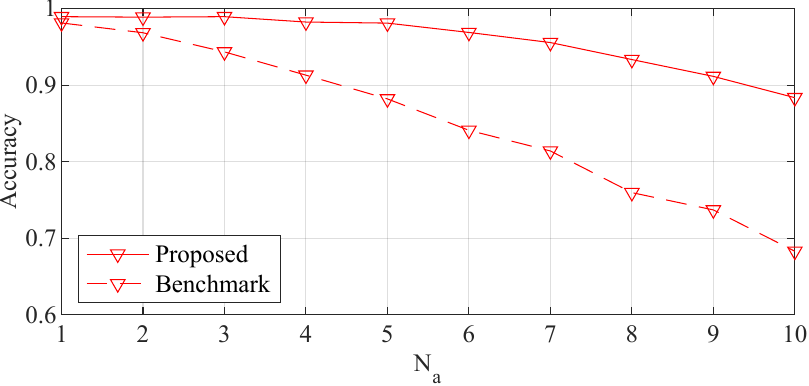}
  \caption{The authentication accuracy versus $N_{a}$ on the simulation test dataset with $T_{\rm rms}=50$ ns, $\text{SNR}=20$~dB, $v_0=1$ m/s, $\Delta t=3$ ms, and $d_{\rm bm}=24$ cm.}
  \label{fig:Acc_threshold}
\end{figure}

Fig.~\ref{fig:Acc_threshold} illustrates the authentication accuracy under varying attack lengths $N_{a}$, where the decision threshold is chosen individually for each authentication scheme so as to maximize its accuracy. The benchmark algorithm performs well for short attack lengths (i.e., $N_{a}=1,2,3$), but degrades rapidly as $N_{a}$ increases, because the reference CSI becomes outdated and decorrelates from Bob's channel at the next legitimate transmission. In contrast, the channel prediction-based scheme sustains markedly better performance at larger $N_{a}$, since the predicted CSI tracks the anticipated legitimate channel evolution across the entire attack interval. 

\section{Conclusion}
In this paper, we studied consecutive spoofing attack in mobile Wi-Fi and proposed a channel prediction-based authentication scheme. By leveraging a Transformer-based channel prediction module to predict the evolution of legitimate CSI during the spoofing interval, the proposed method effectively mitigates the channel decorrelation problem during consecutive spoofing attacks. Simulation results under Rayleigh channel models showed that the proposed approach outperforms the benchmark scheme that does not employ channel prediction, providing robust authentication even under extended attacks.
\bibliographystyle{IEEEtran}
\bibliography{IEEEabrv,cites}

\end{document}